# Broadband EPR Spectroscopy in Diverse Field Conditions Using Optically Detected Nitrogen-Vacancy Centers in Diamond

C. M. Purser[1] *, V. P. Bhallamudi[1,2], C. S. Wolfe[1], H. Yusuf[1], B. A. McCullian[1], C. Jayaprakash[1], M. E. Flatté[3], and P. C. Hammel[1]

[1] Department of Physics, The Ohio State University, Columbus, Ohio, USA
[2] Department of Physics, Indian Institute of Technology Madras, Chennai, India
[3] Department of Physics and Astronomy, The University of Iowa, Iowa City, Iowa, USA

E-mail: purser.6@osu.edu and hammel@physics.osu.edu

## Abstract

Paramagnetic magnetic resonance, a powerful technique for characterizing and identifying chemical targets, is increasingly used for imaging; however, low spin polarization at room temperature and moderate magnetic fields poses challenges for detecting small numbers of spins. In this work, we use fluorescence from nitrogen-vacancy (NV) centers in diamond to detect the electron paramagnetic resonance (EPR) spectrum of optically inactive target spins under various conditions of field magnitude and orientation. The protocol requires neither direct microwave manipulation of the NV spins nor spectral overlap between NV and target spin resonances, thus enabling broadband detection. This unexpected non-resonant coupling is attributable to a two-phonon process that relaxes NV spins proximate to the fluctuating dipole moment of the target spin, suggesting that the sensitivity is determined by the dipole-dipole coupling strength. This approach holds promise for sensitive EPR detection, particularly in settings where control over the diamond-crystal orientation is difficult. This is notably the case for biological sensing applications, where nanodiamonds are being pursued for their bright, stable fluorescence and biocompatibility.

## 1. Introduction

Electron paramagnetic resonance (EPR) provides a spectroscopically precise in situ probe for identifying, locating, and probing the environment of target spins, such as spin labels or oxygen radicals [1]. However, enhancing the application of EPR to enable detection of trace amounts of target spins or to improve the spatial resolution by detecting signals from small volumes is challenging due to the inherent weakness of spin coupling to inductive probes. Low spin polarization at room temperature in moderate applied magnetic fields results in small net magnetic moments, further limiting detection sensitivity in ambient conditions. An alternative to inductive techniques is optically detected magnetic resonance (ODMR) of nitrogen vacancy (NV) defect centers in diamond [2, 3]. ODMR relies on the spin-dependent fluorescence intensity from NV centers. Magnetic resonance (from proximate electron or nuclear spins) can be monitored via their effects on the NV spin-state populations resulting in reduced fluorescence intensity.

Here, we use fluorescence contrast from NV centers to demonstrate broadband EPR spectroscopy on microscale volumes of paramagnetic spins at low–to–moderate applied fields and at room temperature. The insensitivity of the technique to the field orientation relative to the NV axis makes it suitable for applications in which the diamond orientation is difficult to control. These features point to an easily implementable approach to nanoscale EPR in diverse settings.

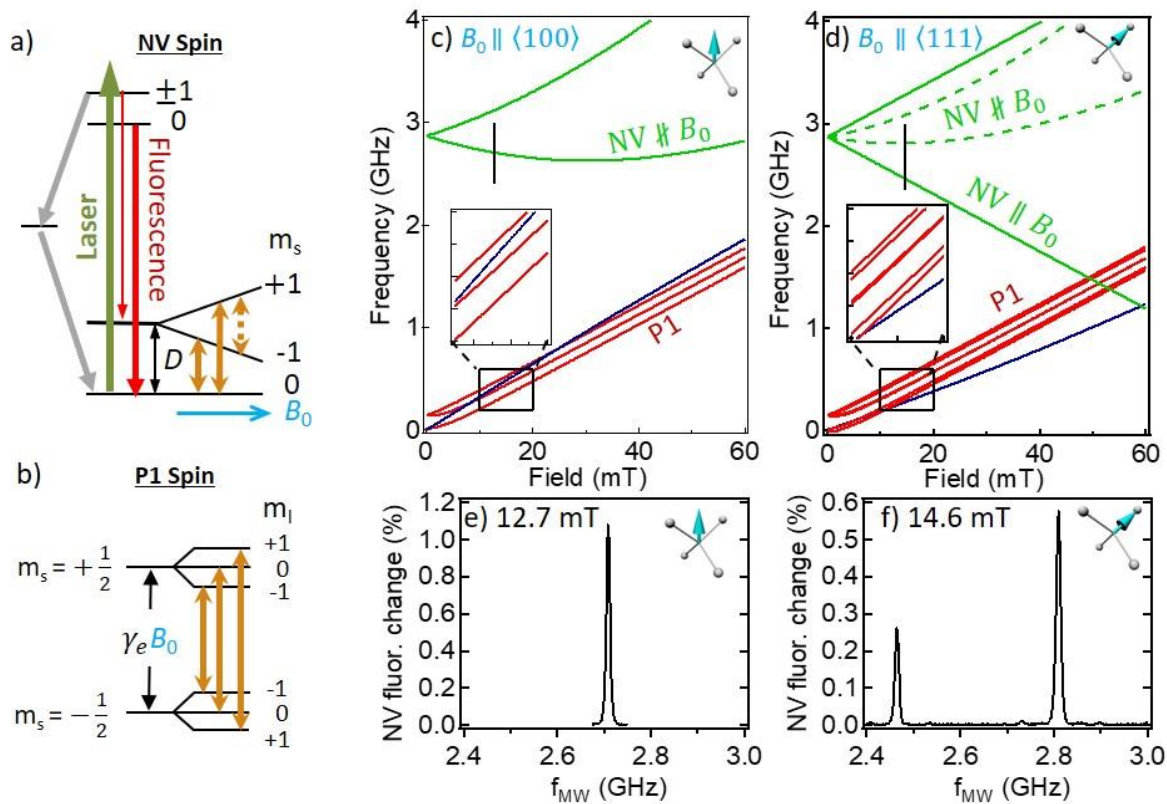

**Figure 1. Energy levels and spin transitions for P1 and NV centers**. Energy diagrams for the **a)** NV sensor spins and **b)** the P1 target spins. **a)** NV ±1 spin states are separated from the $m_s = 0$ state by $D = 2.87$ GHz and Zeeman split from one another in an applied magnetic field. Green laser excites red fluorescence and cyclically polarizes the center into the $m_s = 0$ state through a non-fluorescent intersystem crossing (grey arrows). Transitions from the NV $m_s = 0$ state into the +1 or -1 states results in reduced fluorescence intensity. Double-headed orange arrows indicate transitions driven by microwaves. **b)** P1 electron spin energies are Zeeman split by the external magnetic field and hyperfine-split by interactions with the spin-1 $^{14}$N nucleus. **c)** and **d)** Calculated dispersion of P1 (red), NV (green), and weakly allowed $\Delta m_{mix}=2$ NV (blue) transition frequencies as a function of the applied field for $B_0$ parallel to **c)** $\langle 100 \rangle$ and **d)** $\langle 111 \rangle$ diamond axes. Target P1 spin resonances are spectrally well separated (i.e., non-resonant) from NV transitions at most magnetic fields. **e)** and **f)** Measured ODMR of NV resonances along linecuts indicated in **c)** at 12.7 mT and in **d)** at 14.6 mT, respectively. Lock-in signals register reductions in fluorescence as positive voltages.

Negatively charged NV centers in diamond have been used to achieve very high sensitivity to static and fluctuating magnetic fields at nanometer scales [2, 3]. The sensitivity is attributable in part to high optical polarization, bright spin-dependent fluorescence, and long spin lifetimes of the center. Off-resonant green light (~532 nm) cyclically polarizes the spin-1 center into the $|0\rangle$ spin state and excites red fluorescence (600-800 nm) (**Figure 1 (a)**). The $|0\rangle$ state fluoresces more brightly than the $|+1\rangle$ or $|-1\rangle$ states such that reductions in fluorescence intensity measure even small depletions of the $|0\rangle$ state population. Relatively long spin lifetimes (milliseconds) at room temperature allow interactions with dipole fields from target electron spins that occur on these time scales to modify the NV spin state populations. The NV defect centers can thus be used to optically detect magnetic resonance signals from their surroundings. In addition to the desirable properties of NV centers for spin sensing, diamonds themselves are appealing for bio-applications due to their biocompatibility and their easily cleaned and terminated surfaces. Consequently, NV diamonds, and nanodiamonds in particular, are being actively pursued as in-vivo EPR spectrometers [4-9].

The optically-detected, broadband EPR spectroscopy presented in this work measures changes in the NV fluorescence intensity as a function of the resonant microwave drive frequency and applied field. The target paramagnetic electron spin is associated with optically inactive substitutional nitrogen (P1) centers in diamond [10, 11]. We refer to the sensing protocol as NV-based Non-Resonant Broadband (NV-NRB) detection to highlight the freedom from the need for spectral overlap between the spin transitions for P1 target and NV sensor spins. Absence of spectral overlap is similar to the case of non-resonant coupling seen in NV-ferromagnet systems to detect ferromagnetic resonance [12-14]. However, we postulate a different microscopic mechanism for NV sensitivity to P1 resonances that relies on interaction with the phonon bath, in contrast to the spinwave-based mechanism for the ferromagnetic case.

We find that NV-NRB detection of paramagnetic resonance has several appealing features. Most notably, spectral overlap between the NV and target spins appears to play no role in the coupling. Freedom from this constraint enables broadband detection of target spin spectra, which can be used to precisely measure gyromagnetic ratios and detect anisotropies in the target spin resonance spectrum. Finally, we measure EPR spectra in less than a second of averaging per datum with a relatively straight-forward measurement protocol.

## 2. Methods

The relevant energy levels and spin resonances for the NV sensor and P1 target spins are shown in **Figure 1**. A zero-field splitting ($D$ = 2.87 GHz) between the NV $|0\rangle$ and $|\pm 1\rangle$ states spectrally separates NV transition energies (**Figure 1a**) from those of P1 centers, enabling microwave excitation of P1 transitions without directly driving transitions of the NV spin. The P1 is a spin-½ paramagnetic center with a Landé g-factor of 2. Anisotropic hyperfine coupling to its spin-1 $^{14}$N nucleus splits the P1 resonance into three peaks (**Figure 1b**), where the splitting, $\nu_{hf}$, changes with field orientation [11]. These distinctive spectral features allow signals from P1 centers to be unambiguously identified.

Calculated dispersions for NV and P1 spin transition frequencies as a function of magnetic field strength are shown in **Figures 1c** and **d**. All diamond axes are magnetically equivalent for fields applied along a ⟨100⟩ direction (inset to **Figure 1c**). Thus, microwave excitation of NV-ODMR results in two resonance peaks of equal amplitude, while microwave excitation of P1 resonances results in three peaks of equal amplitude. Fields parallel to ⟨111⟩ are aligned to one of the four possible NV bond axes (inset to **Figure 1d**), and the remaining three axes are magnetically equivalent to one another. The magnetic resonance spectrum for NV centers is then comprised of four transitions. An equal distribution of NV centers along the four crystal axes results in the 1:3:3:1 ratio of signal strengths as a function of frequency. Similarly, for the P1 resonance spectrum, orientation-dependent hyperfine splitting results in five total spectral peaks with relative signal strength 1:3:4:3:1. Further details of the NV and

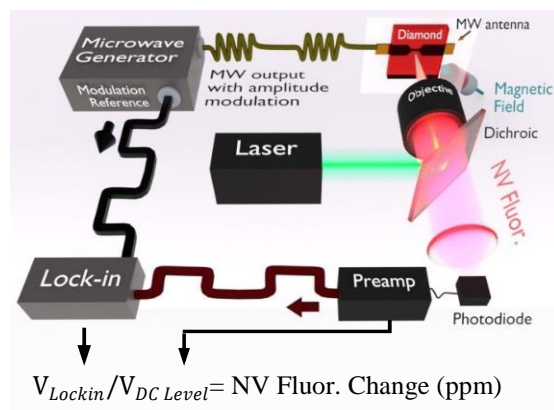

$V_{Lockin}/V_{DC\ Level}$ = NV Fluor. Change (ppm)

**Figure 2. Schematic for NV-NRB detection.** A diamond sample containing NV and P1 centers is mounted on top of a microwave antenna. The static magnetic field needed for resonance can be oriented along a desired crystal axis. Spectra are measured by sweeping the frequency of amplitude modulated microwaves, $f_{MW}$, and measuring reductions in the NV fluorescence intensity using a lock-in, referenced to amplitude modulation of the microwaves. The lock-in voltage (change in fluorescence) is normalized by the pre-amplifier voltage (total fluorescence) and presented as a fractional change in either percent (%) or parts-per-million (ppm).

P1 spin systems and their Hamiltonians that give these resonances are provided in **Supplementary Note 1**.

While we primarily use the microwaves to drive the P1 resonances, we measure the NV-driven ODMR to align the magnetic field and measure the field strength. Experimentally recorded optically detected NV spin resonance peaks are presented in **Figure 1** for at $B_0$ = 12.7 mT along ⟨100⟩ (**e**) and at $B_0$ = 14.6 mT along ⟨111⟩ (**f**).

A schematic for our optically detected broadband EPR spectroscopy is presented in **Figure 2**. A lock-in referenced to the amplitude modulation of microwaves (MW) detects small changes in the intensity of NV fluorescence due to MW drive. Reduced intensity registers as positive voltage signals. Signal strengths are presented as fractional changes by normalizing the lock-in voltage ($V_{Lockin}$) by the total photodiode voltage ($V_{DC\ Level}$) and expressed in parts-per-million (ppm). We record the magnetic resonance spectra either as a function of the frequency of applied microwaves ($f_{MW}$) or as a function of the applied magnetic field ($B_0$). The magnetic field is precisely aligned along either ⟨111⟩ and ⟨100⟩ directions of the NV crystal using a tilt stage. Laser power at the sample was 6 mW and focused through the diamond to a 2 μm spot. Fluorescence was collected from the same objective (x40, NA 0.7) and the excitation laser was filtered out with a dichroic mirror, a 532 nm notch filter, and 620 nm longpass filter.

The single crystal diamond featured here is a (100)-face oriented 1 mm x 1 mm x 0.3 mm high-pressure, high-temperature type 1b single crystal from Sumitomo Electric Inc. containing 50 ppm nitrogen. The diamond was irradiated with 1.5 MeV electrons and annealed at 900 °C in order to produce few ppm NV centers. Based on a continuous Poisson distribution [15], we estimate the average separation between NV and P1 centers to be ~3 nm. Nanodiamond samples are from Adámas Nanotechnologies with average size 100 nm and containing about 500 NV centers each.

## 3. Results

### 3.1 EPR spectroscopy via optical detection of non-resonant NV centers

In this first demonstration of NV-NRB detection of paramagnetic spins, we identify P1 spin resonances by their anisotropic hyperfine splitting in optically detected EPR spectra. **Figure 3** shows measurements of P1 resonance spectra in two field orientations obtained by monitoring changes in the NV fluorescence intensity while sweeping the microwave frequency, $f_{MW}$. The static magnetic field, $B_0$, is oriented along the ⟨100⟩ (panel a) and ⟨111⟩ axes (b), thus changing the anisotropic hyperfine splittings of the P1 centers. Solid red triangles mark the expected peak locations for P1 resonances based on known field strength and orientation as measured from ODMR of NV transitions.

The peak positions as well as the relative peak intensities match well with those reported in literature [4, 6, 11] and calculated values (**Figure 1**). For $B_0 = 12.7$ mT oriented parallel to one of the ⟨100⟩ crystal directions (**Figure 3a**), all the P1 centers are magnetically equivalent, resulting in three P1 peaks of equal intensity. We measure a signal-to-noise ratio of the center P1 peak of approximately 7 using a 100 ms lock-in time-constant and 300 ms lock-in conversion time, for a total measurement time of 300 ms per frequency. In **Figure 3b**, the field is 14.6 mT and aligned along a ⟨111⟩ direction such that there are two magnetically distinguishable directions, resulting in five P1 peaks whose intensities occur at 1:3:4:3:1 ratios, as expected. The lock-in settings were the same as those used in ⟨100⟩-oriented fields, and signals were averaged three times in order to clearly observe all P1 peaks. Furthermore, in **Supplementary Note 2** we demonstrate the ability of this technique to detect P1 spectra at arbitrary field orientations. Though the P1 resonance signals are measured well above the noise, the fluorescence changes are nearly 100 times smaller than ODMR from direct microwave drive of NV resonances (**Figure 1 e** and **f**). Our ability to measure these relatively small changes in fluorescence intensity without long averaging is due to the bright fluorescence from an ensemble of NV centers and to the added sensitivity of a lock-in amplifier to fluorescence changes.

We note that the off-resonant nature of this technique is evident from the dispersion presented in **Figures 1 c** and **d**. The spectral separations between the NV and P1 spins at 12.7 mT and 14.6 mT are more than 2 GHz, much greater than the 10-15 MHz linewidths of either the NV or P1 resonances.

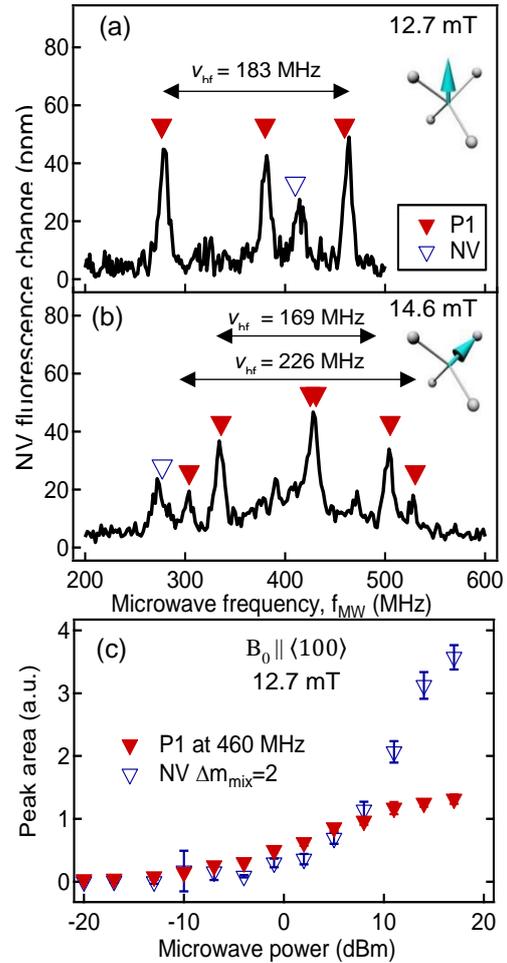

**Figure 3 Optical detection of P1 spectra using NV-NRB.** Representative P1 spectra for applied microwave power at 2 dBm in are shown in **(a)** and **(b)**. Solid red triangles denote the expected resonance peaks for P1 spins, resulting from their hyperfine (hf) structure. Open blue triangles denote the expected transition frequency of the weak, mixed state NV transitions $\Delta m_{mix}$ =2. **(a)** Representative spectrum for $B_0 =$ 12.7 mT along ⟨100⟩ (inset). Three P1 peaks of equal intensity occur in this orientation. **(b)** Representative spectrum for $B_0 =$14.6 mT along ⟨111⟩. P1 peaks with $\nu_{hf}$ =226 MHz have intensity that is one third the intensity of those split by $\nu_{hf}$ =169 MHz due to the 1:3 ratio in non-degenerate populations. **(c)** Microwave power dependences of optical responses from P1 and NV resonances with fields parallel to ⟨100⟩.

We are furthermore able to distinguish P1 resonances from direct MW excitation of normally forbidden transitions (open triangles in **Figure 3**) between NV spin states that are majority $|+1\rangle$ and majority $|-1\rangle$. We refer to these transitions as $\Delta m_{mix} = 2$ [16, 17] because they arise from state mixing due to components of the applied field transverse to the NV quantization axis. NV $\Delta m_{mix} = 2$ transitions can be distinguished from the P1 resonance response through their spectral separation (**Figure 3**) and dispersion (**Figure 4**), as well as from their distinct power dependence (**Figure 3c**). From these data we conclude that the spectral overlap between these weak NV transitions and the P1 resonance play no evident role in the mechanism underlying optical detection of the P1 resonance.

## 3.2 Broadband spectroscopy

The non-resonant coupling between driven P1 centers and optically detected NV centers enables broadband detection of the P1 spectra (**Figure 4**) in several field orientations. Broadband sensing allows us to measure the dispersions of each optically detected resonance feature. Slight differences in the field dependence of the P1 and, for example, $\Delta m_{mix}=2$ NV transitions provides a means of controlling the spectral

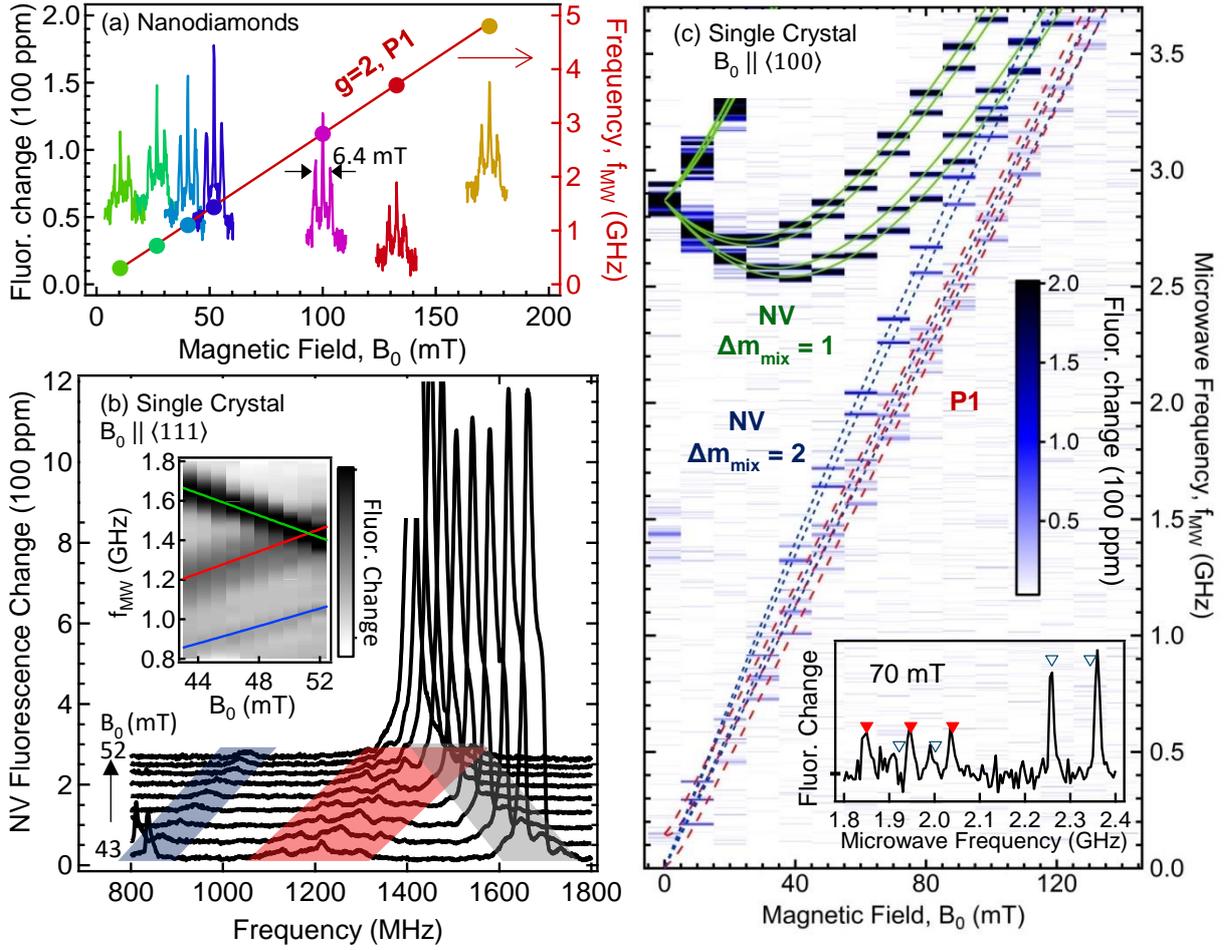

**Figure 4 Demonstration of NV-NRB, broadband detection of P1 spin spectra in various field orientations.** (**a**) Field dependence of optically detected P1 spectra in a powder of nanodiamonds, with randomly oriented NV-axes, at various frequencies: 300 MHz, 715 MHz, 1.1 GHz, 1.435 GHz, 2.8 GHz, 3.7 GHz, 4.8 GHz. The center P1 peaks closely follow the g = 2 dispersion and the spectra have a hyperfine splitting of 6.4 mT in the expected range between 5.7 and 8 mT. (**b**) Spectra collected in a single crystal diamond sample with $\boldsymbol{B_0} \parallel \langle 111 \rangle$. P1 and $\Delta m_{mix} =2$ NV resonances are spectrally well separated for this field orientation and clearly identified. P1 peaks are highlighted in red whereas NV resonances are highlighted in blue. The usual NV peak (between 1400 MHz and 1800 MHz) is much stronger than the optical response from either the P1 resonances or the mixed-state transition. *Inset*: Data are interpolated to highlight P1 dispersions. (**c**) Spectra measured with $B_0$ approximately parallel to $\langle 100 \rangle$. The fractional change is similar to $\boldsymbol{B_0} \parallel \langle 111 \rangle$, demonstrating that NV-NRB technique can even be used with fields transverse to the NV axes. Calculated dispersions for $\Delta m_{mix} =2$ NV transitions are marked in dark blue. P1 dispersions are marked in red. *Inset:* Line-cut at 70 mT with P1 spectral peaks marked with solid red triangles and NV $\Delta m_{mix} =2$ marked with open blue triangles.

separation of these resonances by the external field strength and orientation.

The g = 2 dispersion and the hyperfine triplet structure of P1 spectra are most clearly seen in the measurements on nanodiamonds, presented in **Figure 4a**. The random orientations of the nanodiamonds average the P1 hyperfine splittings over all field directions and result in three peaks. We measure field sweeps for seven microwave frequencies: 300 MHz, 715 MHz, 1.1 GHz, 1.435 GHz, 2.8 GHz, 3.7 GHz, 4.8 GHz. Plotting the frequency of microwave excitation (right axis) as a function of the resonant field of the center P1 peak (solid markers) results in a g=2 dispersion as expected (**Figure 4a**). P1 spectra are measured up to the field limit of our instrumentation, 180 mT. (Variations in the peak intensity with field are most likely due to the frequencydependence of the microwave transmission of the antenna used to excite P1 EPR.)

Single crystal data demonstrate the insensitivity of NV-NRB with respect to the degree of frequency detuning between P1 and NV resonances. In **Figure 4b**, we present data from a single crystal diamond with $B_0||\langle 111 \rangle$. To understand the effect of detuning between the NV and P1 resonances, we focus on the region near the spectral crossing (at 51.2 mT) between P1 resonance and the $|0\rangle \leftrightarrow |-1\rangle$ transition of NV centers with bond axes parallel to $B_0$. Notably, the optical response to the microwave drive of the P1 resonances does not vary as the P1 and NV resonance frequencies approach one another. Furthermore, the $\Delta m_{mix}=2$ NV transitions are spectrally well separated from the P1 resonances in this orientation allowing clearer identification of all the peaks. These data indicate that in our measurements spectral overlap between the P1 and any of the NV ground state transitions are not responsible for the observed changes of NV fluorescence due to driven P1 resonances.

We show the broadband nature of NV-NRB in various field orientations by measuring P1 spectra with $B_0||\langle 100 \rangle$ direction (**Figure 4c**). For fields oriented along $\langle 100 \rangle$, there is no resonance matching between NV $\Delta m_{mix}=1$ and P1 spin transitions. The g = 2 linear P1 dispersion and the $\Delta m_{mix}=2$ dispersions are overlaid in **Figure 4c**, which highlights the detection of P1 resonances over a broad range of fields and frequencies. The inset to **Figure 4c** shows that peaks can be distinguished in linecuts.

### 3.3 Proposed phonon-assisted cross-relaxation mechanism

Understanding the physical mechanism that enables our NV-NRB protocol informs future applications to sensing EPR from spins exterior to the diamond.

This detection method exploits the fact that microwave driven P1 transitions change the steady-state NV spin populations, resulting in reduced NV fluorescence intensity. This is most likely caused by a change in the effective NV longitudinal spin lifetime due to its interaction with driven P1 spins. Normally, a direct spectral overlap between NV and target spins (near 50 mT for fields along a $\langle 111 \rangle$ axis, see **Figure 1d**) is needed to generate fluctuations at the NV resonance frequency and cause spin-relaxation through mutual flip-flops [6, 18]. However, our NV technique functions in the absence of such spectral-matching, enabling broadband sensing. Other broadband NV-based EPR techniques such as double electron-electron resonance (DEER) [19] or dressed state protocols [20] use microwaves to drive both the NV and the P1 spins, and the coupling in these cases may be understood from the effect of P1 spins in the rotating frame of NV spins. In addition, there are detection schemes in which NV centers are sensitive to P1 spins through microwaves at frequencies that compensate for the energy difference between NV and P1 spin resonances [4, 21-23]. However, in this work we only need apply microwaves resonant with the P1 center transitions in order to observe an NV response. A different coupling mechanism must therefore be responsible for the features of our experiments.

We propose that in our experiments the energy difference between the NV and P1 spins (varying from 100 MHz to few GHz depending on the applied field) is provided by phonons. Because gigahertz phonons required for a single-phonon process have vanishingly small density of states in diamond [24, 25], we focus on a two-phonon Raman spin relaxation mechanism. In the two-phonon mechanism, the energy difference between the two phonons, each with energy much higher than gigahertz, will match the difference between the NV and P1 energies. The insensitivity of the signal to both field orientation and the NV-P1 detuning are consistent with this proposed relaxation model.

The strength of the NV-P1 coupling is estimated by taking into account the dipole-dipole interaction as well as the spin-conserving electron-phonon interaction of the centers with the longitudinal acoustic (LA) phonons. Relative to the characteristic rate of a dipole-dipole interaction between the P1 and NV centers, scattering of LA phonons via the deformation potential coupling is essentially instantaneous (effective rates > 1 THz) and can readily provide energy needed for the flip-flop process. The cross-relaxation rate, $\Gamma_x$, will then be determined by the rate-limiting spin-spin dipole interaction in **Equation 1**.

$$\Gamma_x \sim \frac{\mu_0 \mu_B^2}{4\pi r^3 h} \qquad (1)$$

where $\mu_0$ is the vacuum permeability, $\mu_B$ is the magnetic moment of an electron given by the Bohr magneton, $r$ is the separation between NV and P1 spins, and $h$ is Planck's constant. For our average NV-P1 separation of 3 nm, $\Gamma_x \sim 0.5$ MHz.

Having identified a mechanism by which NV and P1 spins may cross-relax, we determine the effect of cross-relaxation on the NV polarization due to microwave excitation of a P1 resonance by solving population rate equations. The ratio of the change in fluorescence due to microwave drive of an NV center resonance versus that due to microwave drive of a P1 transition is about 100. For 1 MHz laser polarization rate, 10 MHz microwave drive strength, and kHz intrinsic spin-lattice relaxation rates for NV and P1 centers [25, 26], we find that a cross-relaxation rate $\Gamma_x \sim 0.1$ MHz would produce this observed ratio (see **Supplementary Note 3**), consistent with the mechanism we have described.

## 4. Discussion

Detection sensitivity is a key parameter for applying this technique to spins external to the diamond. In our experiments, the signals from P1 centers separated from NV centers by ~3 nm reduce fluorescence by less than 100 ppm. Dipole coupling falls off as $1/r^3$ so, regardless of detection method, external target spins must be positioned on diamond having shallowly implanted NV spins. However, spin noise from dangling bonds at the surface can reduce intrinsic NV spin lifetimes $T_{\text{int}}$ (i.e., $T_1$ or $T_2$), thereby reducing sensitivity, which scales as $1/\sqrt{T_{\text{int}}}$ [27, 28]. When applied to shallowly implanted NV centers, our technique, reliant on manipulation of NV $T_1$, should be more robust against degradation by surface conditions than techniques sensitive to $T_2$ [28].

Furthermore, for room temperature applications, $T_1$ is almost independent of the NV concentration [25]. The signal to noise ratio (SNR) can thus be enhanced by a factor of $\sqrt{N_{NV}}$, where $N_{NV}$ is the number of NV centers being measured. While the spatial resolution of an ensemble approach will not reach the single NV limit, it should be better than that of conventional magnetic resonance imaging (MRI), making it attractive for many applications.

In this work we leverage the sensitivity of lock-in detection and NV center fluorescence to acquire P1 spectra (**Figure 3**) with SNR around 7 at modest microwave power and after only 300 ms of averaging per datum. The signal strength may be further improved by optimizing the laser intensity and the microwave power, which was not the focus of this work.

## 5. Conclusion

In summary, the NV-NRB detection of EPR presented here provides an easily implementable nanoscale approach, amenable to room temperature application in a broad field-frequency range regardless of field orientation. This can be expected to greatly enhance the functionality of nanodiamonds that have already attracted much attention as bright, stable biomarkers as well as other biological applications such as in situ thermometry [29, 30].

**Acknowledgements**

We acknowledge support from the Center for Emergent Materials, an NSF MRSEC (Grant DMR-1420451), and the Army Research Office (ARO) (Grant W911NF-16-1-0547). We thank Robert McMichael and Gregory Fuchs for illuminating discussions.

# Supplementary Material:

C. M. Purser[1] *, V. P. Bhallamudi[1,2], C. S. Wolfe[1], H. Yusuf[1], B. A. McCullian[1], C. Jayaprakash[1], M. E. Flatté[3], and P. C. Hammel[1]

**1. Dispersions for NV and P1 centers**

*1.1 NV Dispersion*

The NV center is composed of a nitrogen substitution center adjacent to a carbon vacancy. When the defect is negatively charged, a spin-1 center is formed. In the electronic ground state the spin Hamiltonian is [1]

$$H_{NV} = D \cdot S_z^2 + \gamma \boldsymbol{B} \cdot \boldsymbol{S} + E \cdot (S_x^2 - S_y^2) \tag{S1}$$

where $S$ is the NV electron spin, $D$ = 2.87 GHz is the zero-field crystal field, $\gamma$ = 28.7 GHz/T is the gyromagnetic ratio, and $E \approx$ 3-4 MHz is the strain field. In the absence of an external field, the quantization axis is determined by the zero-field splitting $D$ which generates an effective field along the NV axis. For fields applied parallel to this axis, the $m_s = 0, \pm 1$ basis states remain eigenstates of the Hamiltonian. However, for fields applied transverse to this quantization axis, the eigenstates are mixtures of the axis-quantized basis states. This mixture allows for apparent $\Delta m_{mix} = 2$ transitions.

The energy of the spin splittings is mainly determined by the external magnetic field and the zero-field splitting associated with the NV bond direction. The energy of spin splittings therefore depends on the orientation of the external field relative to the NV bond axis. For a field applied along a $\langle 100 \rangle$ axis, all four possible orientations of the NV center bond have the same angle relative to the external field. This results in two strong microwave (MW) transitions, shown as solid black lines in **Figure S1a**. A third, weaker transition between NV spin eigenstates is due to transverse fields that mix the spin-1 basis states of the NV center. These are shown as a dotted-dashed black line in **Figure S1a**.

However, for magnetic fields aligned along a $\langle 111 \rangle$ crystal axis, one of the NV bond directions aligns with the field, such that spin eigenstates correspond to spin basis states. Because NV centers that are aligned to the external field experience no state mixing due to transverse fields, only two transitions between the $m_s = 0$ and 1 and between the $m_s = 0$ and -1 are possible. These appear as straight, solid black lines in **Figure S1b.** The remaining three NV bond directions have the same 109.5 deg orientation to the external field, resulting in two additional strong MW transitions, as well as one weakly fluorescing transition due to transverse field mixing. These are shown as two curved dashed black lines and a dotted-dashed black line in **Figure S1b.**

*1.1 P1 Dispersion*

The P1 center is composed of a nitrogen substitution center and forms a spin-1/2 center. The Hamiltonian is as follows [1]:

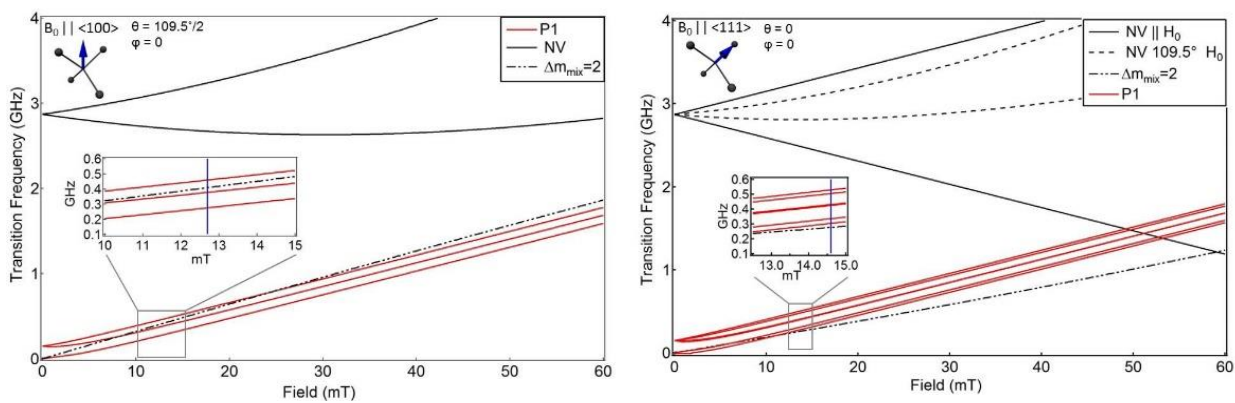

**Figure S1** P1 and NV dispersions as a function of field applied along the a) $\langle \mathbf{100} \rangle$ axis and b) $\langle \mathbf{111} \rangle$ axis.

$$H_{P1} = \gamma \boldsymbol{B} \cdot \boldsymbol{S_P} + A_\parallel I_z S_{P,z} + A_\perp \left(I_x S_{P,x} + I_y S_{P,y}\right) + \gamma_N \boldsymbol{B} \cdot \boldsymbol{I} + Q(I_z^2 - \boldsymbol{I}^2/3) \qquad (S2)$$

where $S_P$ is the P1 electron spin and $\boldsymbol{I}$ is the nuclear spin (spin-1) of the nitrogen atom with corresponding gyromagnetic ratio, $\gamma_N = -3.08$ MHz/T for $^{14}$N. The $^{14}$N isotope occurs in natural abundance (> 99%) for the nitrogen in these diamond crystals. The anisotropic hyperfine coupling between the electron and $^{14}$N nuclei is $A_\parallel$=114 MHz and $A_\perp$ = 81.3 MHz [1]. Electronic transitions are those that preserve nuclear spin (**Figure 1** in the main text), allowing for three distinct resonances for P1 centers of a given orientation. The anisotropy is associated with the orientation of the external field relative to the diamond lattice.

Similar to the NV resonances, for fields aligned along the ⟨100⟩ crystal axis, all four bonds at the same angle to the external field, and the hyperfine splittings for each P1 center are the same. This results in three P1 transitions of equal intensity, whose dispersions are shown in **Figure S1a** as three red solid lines.

For fields orientated along a ⟨111⟩ crystal axis, one fourth of the P1 centers are oriented along the field direction while the remaining P1 centers are 109.5 deg to the field. These different orientations result in two sets of hyperfine splittings producing a total number of five P1 spectral peaks with ratios of relative intensity 1:3:4:3:1. The dispersions for the P1 transitions are shown in **Figure S1b** as five solid red lines.

## 2. Frequency sweep spectra at an arbitrary field orientation

Data presented in the main paper use fields oriented along either the ⟨111⟩ or ⟨100⟩ crystal directions in order to demonstrate the applicability of our technique to different field orientations. However, the field need not be oriented along one of these directions in order to detect the P1 resonances. In **Figure S2** the field is not purposely aligned to any crystal direction, but from the NV ODMR spectrum we find that the field is 7.9 mT with angles θ = 43.8 deg and φ = 5.1 deg , where θ is the angle from an NV axis in a plane defined by one other crystal bond and φ defines a rotation about the original NV bond axis (consistent with what has been shown in the insets for **Figure S1**). Red diamonds mark the calculated positions of P1 transitions and open dark blue triangles mark the calculated positions of NV $\Delta m_{\text{mix}} = 2$ transitions, though the calculated NV transitions at 272 MHz and 182 MHz do not seem to correspond to any spectral features.

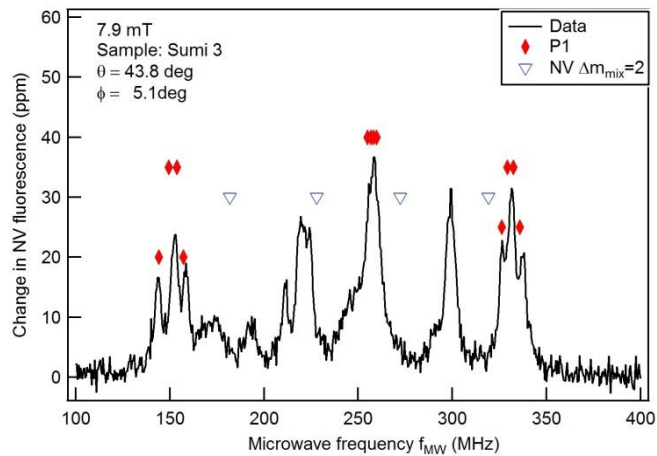

**Figure S2** NV fluorescence change plotted as a function of frequency sweep with a 7.9 mT at angles θ = 5.1 deg and φ = 43.8 deg. The calculated P1 resonances for this field strength and orientation are shown in red solid diamonds. The calculated $\Delta m_{\text{mix}} = 2$ NV transition frequency is marked by an open dark blue triangle. The feature at 300 MHz has no obvious origin (including harmonics from the MW generator that can sometime excite NV ground state transitions).

## 3. Calculating cross-relaxation rates

In the main text, we have described a cross-relaxation mechanism that enables polarization transfer between the NV and P1 spin populations. Here we calculate the expected steady-state change in NV populations in the presence of cross-relaxation with P1 centers under two different microwave drive conditions: 1. MW drive of P1 resonances and 2. MW drive of an NV resonance. The NV polarization We compare these with the observed relative change in order to estimate the cross-relaxation strength. A diagram of the spin-levels and transition process are shown in **Figure S3.**

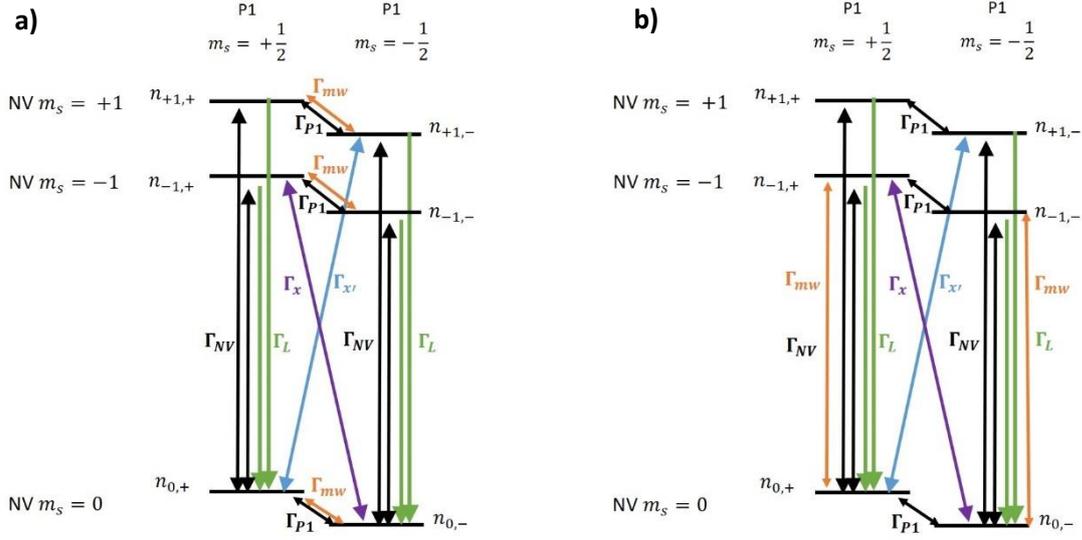

**Figure S3** Diagrams of the coupled NV-P1 spin states with relevant transitions for the case where a) the microwaves are resonant with the P1 transition b) the microwaves are resonant with an NV transition.

The NV polarization is defined as $p_{NV} = (n_{0,-} + n_{0,+}) - (n_{-1,-} + n_{-1,+} + n_{+1,-} + n_{+1,+})$, and the P1 polarization would be defined as $p_{P1} = (n_{0,-} + n_{-1,-} + n_{+1,-}) - (n_{0,+} + n_{-1,+} + n_{+1,+})$.

### 3.1 MW drive of P1 centers

$$\frac{dn_{0,-}}{dt} = -\Gamma_{sp}\,\delta n_{0,-} + \Gamma_L(n_{-1,-} + n_{+1,-}) - \Gamma_{mw}(n_{0,-} - n_{0,+}) - \Gamma_{x\prime}(\delta n_{0,-} - \delta n_{-1,+})$$

$$\frac{dn_{0,+}}{dt} = -\Gamma_{sp}\delta n_{0,+} + \Gamma_L(n_{-1,+} + n_{+1,+}) + \Gamma_{mw}(n_{0,-} - n_{0,+}) - \Gamma_x(\delta n_{0,+} - \delta n_{+1,-})$$

$$\frac{dn_{-1,-}}{dt} = -\Gamma_{sp}\delta n_{-1,-} - \Gamma_L(n_{-1,-}) - \Gamma_{mw}(n_{-1,-} - n_{-1,+})$$

$$\frac{dn_{-1,+}}{dt} = -\Gamma_{sp}\delta n_{-1,+} - \Gamma_L(n_{-1,+}) + \Gamma_{mw}(n_{-1,-} - n_{-1,+}) + \Gamma_{x\prime}(\delta n_{0,-} - \delta n_{-1,+})$$

$$\frac{dn_{+1,-}}{dt} = -\Gamma_{sp}\delta n_{+1,-} + \Gamma_L(n_{+1,-}) - \Gamma_{mw}(n_{+1,-} - n_{+1,+}) + \Gamma_x(\delta n_{0,+} - \delta n_{+1,-})$$

$$\frac{dn_{+1,+}}{dt} = -\Gamma_{sp}\delta n_{+1,+} - \Gamma_L(n_{+1,+}) + \Gamma_{mw}(n_{+1,-} - n_{+1,+})$$

where the $\delta n's$ denote the deviation, $n - n_{th}$, of the populations from thermal equilibrium (i.e., $n = n_{th}$ in the absence of microwave drive or laser polarization). The various rates of interaction are defined in **Table S1.** The cross-relaxation rates ($\Gamma_{x\prime}$ and $\Gamma_x$) are defined separately, though we expect that they will be equal. The microwave transition rates define transitions between NV $m_s = 0$ and $m_s = -1$ spin states that do not affect the P1 spin states.

**Table S1** Rate constants for transitions between different NV-P1 spin states.

| | | |
|---|---|---|
| $\Gamma_{sp}$ | ~ 2 kHz | Spin-lattice relaxation rate for NV and P1 centers |
| $\Gamma_L$ | ~ 1 MHz | Laser polarization rate |
| $\Gamma_{mw}$ | 0-20 MHz | MW transition rate |
| $\Gamma_x$ | Expected: ~0.5 MHz | Cross-relaxation rate $n_{0,+} \leftrightarrow n_{+1,-}$ |
| $\Gamma_{x\prime}$ | Expected: ~0.5 MHz | Cross-relaxation rate $n_{0,-} \leftrightarrow n_{-1,+}$ |

### 3.2 MW drive of NV centers

We calculate the expected change in NV populations due to microwave drive of NV (**Figure S3b**), transitions between NV $m_s = 0$ and $m_s = -1$ spin states that conserve P1 spin using the following rate equations:

$$\frac{dn_{0,-}}{dt} = -\Gamma_{sp}\delta n_{0,-} + \Gamma_L(n_{-1,-} + n_{+1,-}) - \Gamma_{mw}(n_{0,-} - n_{-1,-}) - \Gamma_{x\prime}(\delta n_{0,-} - \delta n_{-1,+})$$

$$\frac{dn_{0,+}}{dt} = -\Gamma_{sp}\delta n_{0,+} + \Gamma_L(n_{-1,+} + n_{+1,+}) - \Gamma_{mw}(n_{0,+} - n_{-1,+}) - \Gamma_x(\delta n_{0,+} - \delta n_{+1,-})$$

$$\frac{dn_{-1,-}}{dt} = -\Gamma_{sp}\delta n_{-1,-} - \Gamma_L(n_{-1,-}) + \Gamma_{mw}(n_{0,-} - n_{-1,-})$$

$$\frac{dn_{-1,+}}{dt} = -\Gamma_{sp}\delta n_{-1,+} - \Gamma_L(n_{-1,+}) + \Gamma_{mw}(n_{0,+} - n_{-1,+}) + \Gamma_{x\prime}(\delta n_{0,-} - \delta n_{-1,+})$$

$$\frac{dn_{+1,-}}{dt} = -\Gamma_{sp}\delta n_{+1,-} + \Gamma_L(n_{+1,-}) + \Gamma_x(\delta n_{0,+} - \delta n_{+1,-})$$

$$\frac{dn_{+1,+}}{dt} = -\Gamma_{sp}\delta n_{+1,+} - \Gamma_L(n_{+1,+})$$

### 3.3 Comparison and estimated $\Gamma_x$

Solving for the steady state populations above, we can then solve for the NV polarization in terms of the rate constants. The change in NV polarization, $p_{NV}(\Gamma_{mw} = 0) - p_{NV}(\Gamma_{mw} \neq 0)$, are plotted in **Figure S4a** for $\Gamma_{sp} = 2$ kHz, $\Gamma_L = 1$ MHz, $\Gamma_x = 0.1$ MHz, and $\Gamma_{x\prime} = 0.1$ MHz. Because we compare the relative fluorescence change from MW drive on NV resonance $\Delta p_{NV}(f_{MW} = f_{NV})$ to the fluorescence change from MW drive of a P1 resonance $\Delta p_{NV}(f_{MW} = f_{P1})$, it is useful to plot this ratio as a function of the cross-relaxation rate **Figure S4b.** For a ratio of NV polarization change ~100, the cross-relaxation rate is expected to be around 0.1 MHz, irrespective of laser polarization rate.

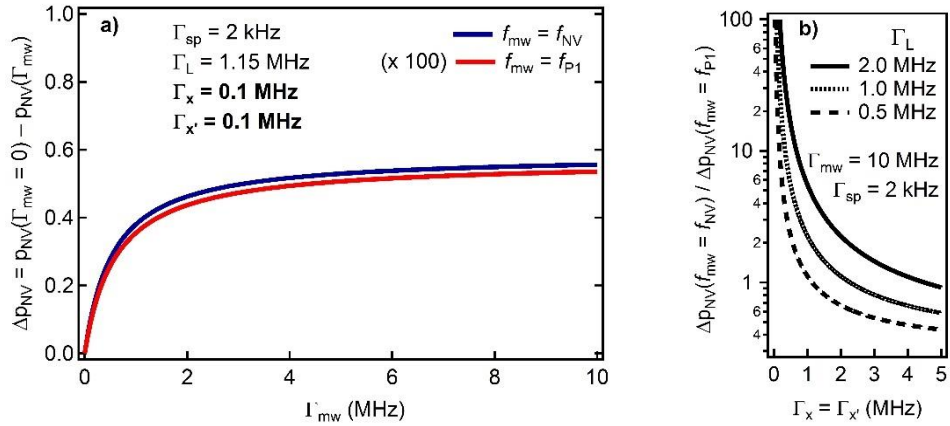

**Figure S4** Resulting calculation for the steady-state change in NV polarization as a result of MW drive resonant with an NV transition (blue) and with a P1 transition (red). **a)** Change in NV polarization plotted as a function of MW drive strength. MW drive of P1 resonances (red) is multiplied by a factor of 100 to demonstrate similar MW power dependence and x100 difference in the change in NV populations. **b)** Ratio of the change in NV polarization with microwaves driving NV transition versus P1 transition plotted as a function of cross-relaxation strength. The MW drive is at 10 MHz and ratio of NV polarization changes are plotted for three laser polarization rates. A ratio of about 100, consistent with the measured fluorescence changes, is achieved for $\Gamma_x = \Gamma_{x\prime} \sim 0.1$ MHz irrespective of laser polarization rate.